\documentclass[12pt]{iopart}

\usepackage{graphicx}
\begin{document}

\title {Cryptanalysis and improvement of the quantum private comparison protocol based on Bell entangled states}

\author{Wen-Jie Liu$^{1,2}$, Chao Liu$^{2}$, Yu Zheng$^{1,2}$ and Zheng-Fei Chen$^{2}$}

\address{
$^{1}$Jiangsu Engineering Center of Network Monitoring, Nanjing University of Information Science \& Technology, Nanjing 210044, People's Republic of China\\
$^{2}$School of Computer and Software, Nanjing University of Information Science \& Technology, Nanjing 210044, People's Republic of China}
\ead{wenjieliu1106@gmail.com}

\begin{abstract}
Recently, Liu et al. [Commun. Theor. Phys. 57, 583, 2012] proposed a quantum private comparison protocol based on entanglement swapping of Bell states, which aims to securely compare the equality of two participants' information with the help of a semi-honest third party (TP). However, this study points out there is a fatal loophole in this protocol, i.e., TP can obtain all of the two participants secret inputs without being detected through making a specific Bell-basis measurement. To fix the problem, a simple solution, which uses one-time eavesdropper checking with decoy photons instead of twice eavesdropper checking with Bell states, is demonstrated. Compared with the original protocol, it also reduces the Bell states consumption and simplifies the steps in the protocol.

\end{abstract}

\maketitle

\section{Introduction}
Since Bennett and Brassard [1] proposed the first quantum key distribution protocol in 1984, various quantum cryptography protocols have been flourished by utilizing quantum mechanics principles, including quantum secret sharing (QSS) [2-4], quantum key distribution (QKD) [5-7], quantum teleportation [8, 9] and quantum direct communication (QDC) [10-13], etc. The main purpose is to provide unconditionally secure information exchange on basis of the law of quantum mechanics.

As a fundamental primitive in modern cryptography, secure multiparty computation (SMC) has been a research hotspot in classical cryptography. It originated in the millionaire problem introduced by Yao [14], in which two millionaires hope to determine who is richer without revealing the precise amount of their fortunes. As a first step to solve the millionaire problem, private comparison of equality was proposed by Boudot et al. [15] to compare the equality of two millionaires' property, without disclosing any actual information. However, the security of SMC is based on the computational complexity assumptions, so it may be seriously threatened by the powerful quantum computer. Fortunately, quantum cryptography, which is regarded as one of the most promising applications of quantum mechanics, is able to guarantee the unconditional security of the information.

As the quantum counterpart of private comparison of equality, quantum private comparison of equality (QPCE) has become an important branch of quantum cryptography. QPCE allows two participants to privately compare the equality of their secret information based on the properties of quantum mechanics, without disclosing any information about their secrets. However, Lo [16] pointed out that the equality function cannot be securely evaluated by two-party protocol in quantum scenario. Therefore, an additional condition with a third party (TP) is necessary to reach the goal of private comparison. The pioneering QPCE protocol was proposed by Yang et al. [17] in 2009, in which a one-way hash function was used to calculate the hash values of two participants' secret inputs firstly, and then these hash values were encoded into the photons of EPR pairs. In essence, its security is guaranteed by the hash function. Since then, with different categories of quantum states, many other QPCE protocols have been proposed [18-24]. Recently, Liu et al. [25] proposed a QPCE protocol based on Bell states. In the protocol, the characteristic of quantum entanglement swapping is utilized to realize the comparison task, and TP is assumed as a semi-honest third party. However, while revisiting Liu et al.'s protocol, we find that it has a fatal security loophole, i.e., TP can obtain all of the two participants¡¯ secret information without being detected through making a specific Bell-basis measurement (we called it as TP's measurement attack), which is obviously against the QPCE's requirements [26]. Furthermore, a simple and efficient solution is given herein to eliminate this security loophole.

The rest of this paper is organized as follows. In Sect. 2, Liu et al.'s protocol is briefly reviewed. In Sect. 3, the security loophole is analyzed and an improved strategy is given. Finally, a short conclusion is given in Sect. 4.

\section{Review of Liu et al.'s Protocol}
In Ref. [25], Liu et al. proposed a QPCE protocol based on Bell states $|\Phi^{\pm}\rangle=\frac{1}{\sqrt{2}}(|00\rangle)\pm|11\rangle)$, $|\Psi^{\pm}\rangle=\frac{1}{\sqrt{2}}(|01\rangle)\pm|10\rangle)$. In the protocol, all the three parties need to prepare Bell states, and the private comparison task is fulfilled by utilizing the entanglement swapping between the Bell states of the participants' and TP's. The main procedures of the protocol are as follows.

\textbf{Prerequisite.} Alice and Bob agree that the states $|\Phi^{+}\rangle$, $|\Phi^{-}\rangle$, $|\Psi^{+}\rangle$, $|\Psi^{-}\rangle$ represent the information `00', `01', `10', `11', respectively.

\textbf{Step 1.} Alice (Bob) equally divides her (his) binary representation of secret inputs $X$ ($Y$) into $\lceil L/2\rceil$ groups, and they are denoted by $G^{A}_{1},G^{A}_{2},...,G^{A}_{\lceil L/2\rceil}$ $(G^{B}_{1},G^{B}_{2},...,G^{B}_{\lceil L/2\rceil})$.

\textbf{Step 2.} Alice, Bob and TP prepare $\lceil L/2\rceil$ Bell states in $|\Phi^{+}\rangle$, and take the first (second) particle from each state to form their own sequences $S^{A}_{1}(S^{A}_{2})$, $S^{B}_{1}(S^{B}_{2})$ and $S^{C}_{1}(S^{C}_{2})$, respectively.

\textbf{Step 3.} Alice and TP prepare an order $L^{'}$-length Bell states sequences in $|\Phi^{+}\rangle$ once again, which are denoted by $T_{A^{'}}$ $(T_{C^{'}})$ sequences, respectively. $T_{A^{'}}$ and $T_{C^{'}}$ will be used to perform eavesdropper checking. Then Alice (TP) inserts the first and second particles of every Bell state in sequence $T_{A^{'}}$ $(T_{C^{'}})$ into sequences $S^{A}_{1}(S^{C}_{1})$ and $S^{A}_{2}(S^{C}_{2})$ at the same positions, respectively, and get $S^{A^{'}}_{1}(S^{C^{'}}_{1})$ and $S^{A^{'}}_{2}(S^{C^{'}}_{2})$. Afterwards, they exchange $S^{A^{'}}_{2}$ and $S^{C^{'}}_{2}$ (shown in Fig. 1a). Then Alice and TP perform the eavesdropper checking of Alice-TP (TP-Alice) quantum channel by using the correlation of Bell states in $T_{A^{'}}$ $(T_{C^{'}})$. If there is no eavesdropper, Alice and TP discard the checking particles, and continue the next step.

\textbf{Step 4.} Alice performs the Bell-basis measurement on corresponding two particles in $S^{A}_{1}$, $S^{C}_{2}$, and the measurement outcomes is denoted by $M^{A}_{j}$, and she calculates $R^{A}_{j}$ in terms of $M^{A}_{j}$. If $M^{A}_{j}$ is $|\Phi^{+}\rangle$ ($|\Phi^{-}\rangle$, $|\Psi^{+}\rangle$, $|\Psi^{-}\rangle$), then $R^{A}_{j}$ is `00'(`01', `10', `11'). As a result, the corresponding two particles in $S^{C}_{1}$, $S^{A}_{2}$ on TP's hand will be collapsed into one of four Bell states (shown in Fig. 1b). The new sequences are denoted by $S^{C^{''}}_{1}$, $S^{C^{''}}_{2}$.

\textbf{Step 5.} Bob and TP prepare another order $L^{'}$-length Bell states sequences in $|\Phi^{+}\rangle$, and insert them into $S^{B}_{1}$, $S^{B}_{2}$ and $S^{C^{''}}_{1}$, $S^{C^{''}}_{2}$, respectively, which is same as that in Step 3. After that, Bob and TP exchange the second particles sequence (shown in Fig. 1c), and perform the eavesdropper checking in the same way. If the quantum channels of Bob-TP and TP-Bob are secure, Bob and TP discard the checking particles, and continue the next step.

\textbf{Step 6.} Bob performs the Bell-basis measurement on the corresponding two particles in $S^{B}_{1}$, $S^{C^{''}}_{2}$ on his hand (the outcome is denoted by $M^{B}_{j}$), then the corresponding two particles in $S^{C}_{1}$, $S^{A}_{2}$ will be collapsed into one of four Bell states (shown in Fig. 1d). At the same time, TP performs the same Bell-basis measurement on the particles in $S^{C^{''}}_{1}$, $S^{B}_{2}$, and gets the outcome $M^{C}_{j}$. Then, Bob (TP) can calculate $R^{B}_{j}$ $(R^{C}_{j})$ in terms of $M^{B}_{j}$ $(M^{C}_{j})$. If $M^{B}_{j}$ $(M^{C}_{j})$ is $|\Phi^{+}\rangle$ ($|\Phi^{-}\rangle$, $|\Psi^{+}\rangle$, $|\Psi^{-}\rangle$), then $R^{B}_{j}$ $(R^{C}_{j})$ is `00'(`01', `10', `11'). For the convenience of calculation, $R^{C}_{j}$ can be represented as below, $R^{C}_{j}=(r^{C1}_{j}r^{C2}_{j})$.

\textbf{Step 7.} Alice and Bob calculate $R_{j}=(R^{A}_{j}\oplus G^{A}_{j})\oplus (R^{B}_{j}\oplus G^{B}_{j})=(r^{1}_{j}r^{2}_{j})$ $(i\leq j\leq \lceil L/2\rceil)$, and send the result $R_{j}$ to TP. TP calculates $R=\sum^{\lceil L/2\rceil}_{j=1}((r^{1}_{j}\oplus r^{C1}_{1})+(r^{2}_{j}\oplus r^{C2}_{1}))$.

\textbf{Step 8.} TP sends $R$ to Alice and Bob, if $R=0$, then $X=Y$; otherwise $X\neq Y$.

\begin{figure}
  \centering
    \includegraphics[width=0.75\textwidth]{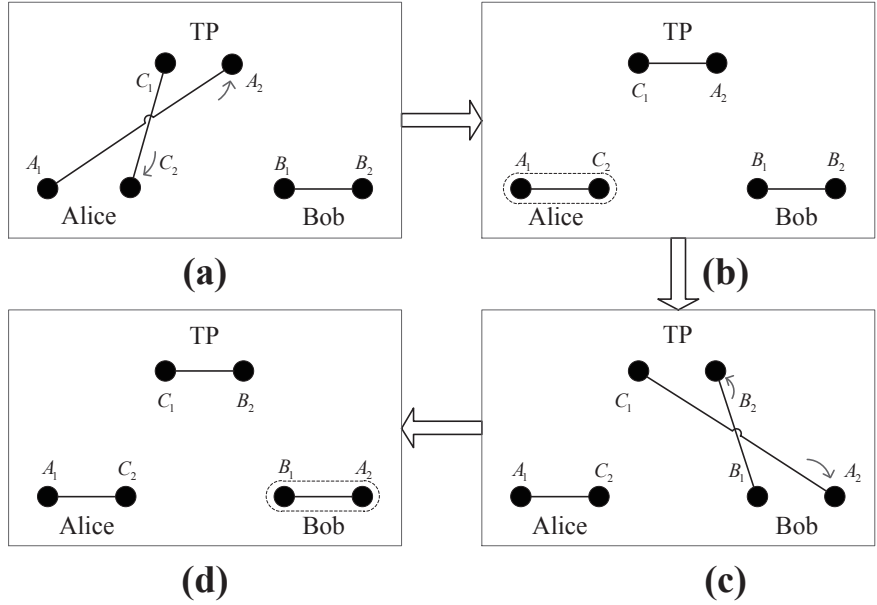}
  \centering
\caption{The entanglement swapping of Bell states among the three parties in Liu et al.'s protocol. \textbf{(a)} Alice and TP exchange their second particles ($A_{2}$, $C_{2}$) of Bell states. \textbf{(b)} Particles $C_{1}$ and $A_{2}$ on TP's hand become entangled together after Alice performs the Bell-basis measurement on particles $A_{1}$ and $C_{2}$. \textbf{(c)} TP and Bob continue to exchange particles $A_{2}$ and $B_{2}$. \textbf{(d)} Particles $C_{1}$ and $B_{2}$ on TP's hand form a new Bell entangled state after Bob performs the Bell-basis measurement on particles $B_{1}$ and $A_{2}$.}
\label{fig:1}
\end{figure}

As shown above, the eavesdropper checking of Alice-TP and TP-Bob quantum channels are performed step by step, that means it needs to make twice eavesdropper checking to confirm the secure of the whole quantum channels from Alice to Bob. And in the protocol, TP is assumed to be semi-honest third party [27], who is allowed to misbehave on its own but cannot conspire with either of two participants. So, TP can make any attacks in the communication process as long as he does not collude with the two participants. That is to say, TP is likely to launch the measurement attack to obtain the participants' secret inputs. The more details will be described in Sect. 3.

\section{TP's Measurement Attack and the Improvement}
\subsection{TP's Measurement Attack on Liu et al.'s Protocol}
While reexamining Liu et al.'s protocol, we find that there exists a security loophole in it. TP could launch the measurement attack to obtain all the two participants' secret inputs without being detected. The detailed analysis is as follows.

As depicted in Step 4, after Alice performs Bell-basis measurement on two particles in $S^{A}_{1}$, $S^{C}_{2}$, the corresponding two particles on TP's hand will be collapsed into one of four Bell states. Then the correspondence of particles after the entanglement swapping can be shown as below,
\begin{eqnarray}
\fl |\Phi^{+}\rangle_{A_{1}A_{2}}\oplus |\Phi^{+}\rangle_{C_{1}C_{2}} =\frac{1}{\sqrt{2}}(|00\rangle+|11\rangle)_{A_{1}A_{2}}\oplus (|00\rangle+|11\rangle)_{C_{1}C_{2}} \nonumber\\
=\frac{1}{2}(|00\rangle_{A_{1}C_{2}}|00\rangle_{C_{1}A_{2}}+|01\rangle_{A_{1}C_{2}}|10\rangle_{C_{1}A_{2}} \nonumber\\
+|10\rangle_{A_{1}C_{2}}|01\rangle_{C_{1}A_{2}}+|11\rangle_{A_{1}C_{2}}|11\rangle_{C_{1}A_{2}}) \nonumber\\
=\frac{1}{2}(|\Phi^{+}\rangle_{A_{1}C_{2}}|\Phi^{+}\rangle_{C_{1}A_{2}}+|\Phi^{-}\rangle_{A_{1}C_{2}}|\Phi^{-}\rangle_{C_{1}A_{2}} \nonumber\\
+|\Psi^{+}\rangle_{A_{1}C_{2}}|\Psi^{+}\rangle_{C_{1}A_{2}}-|\Psi^{-}\rangle_{A_{1}C_{2}}|\Psi^{-}\rangle_{C_{1}A_{2}})
\end{eqnarray}
From the above Eq. 1, we can know that there is a direct relationship between Alice's and TP's Bell states, i.e., they are always the same. For example, if the outcome of Bell measurement on the pair on TP's hand $(C_{1}, A_{2})$ is $|\Phi^{+}\rangle$, the Bell state on Alice's hand $(A_{1}, C_{2})$ must be $|\Phi^{+}\rangle$. Then, TP can steal Alice's and Bob's secrets through the below approach.

Suppose TP wants to steal Alice's secrets. TP makes the Bell-basis measurement on the particles on his hand in Step 4, and gets the outcome $M^{C^{'}}_{j}$. Since Alice's and TP's Bell states are the same, TP can deduce Alice's corresponding outcome, $M^{A}_{j}=M^{C^{'}}_{j}$, and then gets its binary value $R^{A}_{j}$. More seriously, this behavior will not be detected since it will not affect the states on Alice's hand. On the other hand, Alice and Bob need to calculate $R_{j}=E^{A}_{j}\oplus E^{B}_{j}$ $(E^{A}_{j}=R^{A}_{j}\oplus G^{A}_{j}, E^{B}_{j}=R^{B}_{j}\oplus G^{B}_{j})$ cooperatively in Step 7, that means, the encrypted messages $E^{A}_{j}$, $E^{B}_{j}$ are required to be sent through the public channel. As we all known, the classical information transmitted through the public channel can be arbitrarily obtained without being detected. Therefore, TP can get $E^{A}_{j}$ without being detected. In terms of $E^{A}_{j}$ and $R^{A}_{j}$, TP can calculate Alice's input $G^{A}_{j}=E^{A}_{j}\oplus R^{A}_{j}$. For $j=1$ to $\lceil L/2\rceil$, TP can easily get all of Alice's secret inputs $X$.

TP can also steal Bob's secrets. As depicted in Step 6, Bob and TP perform the Bell-basis measurement on the corresponding two particles on their hands, respectively, which will result in the entanglement swapping between TP and Bob. And the correspondence of particles can be one of the following four equations,
\begin{eqnarray}
\fl |\Phi^{+}\rangle_{C_{1}A_{2}}\oplus |\Phi^{+}\rangle_{B_{1}B_{2}}
=\frac{1}{2}(|\Phi^{+}\rangle_{C_{1}B_{2}}|\Phi^{+}\rangle_{B_{1}A_{2}}+|\Phi^{-}\rangle_{C_{1}B_{2}}|\Phi^{-}\rangle_{B_{1}A_{2}} \nonumber\\
+|\Psi^{+}\rangle_{C_{1}B_{2}}|\Psi^{+}\rangle_{B_{1}A_{2}}-|\Psi^{-}\rangle_{C_{1}B_{2}}|\Psi^{-}\rangle_{B_{1}A_{2}})
\end{eqnarray}
\begin{eqnarray}
\fl |\Phi^{-}\rangle_{C_{1}A_{2}}\oplus |\Phi^{+}\rangle_{B_{1}B_{2}}
=\frac{1}{2}(|\Phi^{+}\rangle_{C_{1}B_{2}}|\Phi^{-}\rangle_{B_{1}A_{2}}+|\Phi^{-}\rangle_{C_{1}B_{2}}|\Phi^{+}\rangle_{B_{1}A_{2}} \nonumber\\
|\Psi^{+}\rangle_{C_{1}B_{2}}|\Psi^{-}\rangle_{B_{1}A_{2}}+|\Psi^{-}\rangle_{C_{1}B_{2}}|\Psi^{+}\rangle_{B_{1}A_{2}})
\end{eqnarray}
\begin{eqnarray}
\fl |\Psi^{+}\rangle_{C_{1}A_{2}}\oplus |\Phi^{+}\rangle_{B_{1}B_{2}}
=\frac{1}{2}(|\Phi^{+}\rangle_{C_{1}B_{2}}|\Psi^{+}\rangle_{B_{1}A_{2}}+|\Phi^{-}\rangle_{C_{1}B_{2}}|\Psi^{-}\rangle_{B_{1}A_{2}} \nonumber\\
+|\Psi^{+}\rangle_{C_{1}B_{2}}|\Phi^{+}\rangle_{B_{1}A_{2}}-|\Psi^{-}\rangle_{C_{1}B_{2}}|\Phi^{-}\rangle_{B_{1}A_{2}})
\end{eqnarray}
\begin{eqnarray}
\fl |\Psi^{-}\rangle_{C_{1}A_{2}}\oplus |\Phi^{+}\rangle_{B_{1}B_{2}}
=\frac{1}{2}(|\Phi^{+}\rangle_{C_{1}B_{2}}|\Psi^{-}\rangle_{B_{1}A_{2}}+|\Phi^{-}\rangle_{C_{1}B_{2}}|\Psi^{+}\rangle_{B_{1}A_{2}} \nonumber\\
|\Psi^{+}\rangle_{C_{1}B_{2}}|\Phi^{-}\rangle_{B_{1}A_{2}}+|\Psi^{-}\rangle_{C_{1}B_{2}}|\Phi^{+}\rangle_{B_{1}A_{2}})
\end{eqnarray}
Since TP had made the Bell-basis measurement on particles $C_{1}$ and $A_{2}$, and got the outcome $M^{C^{'}}_{j}$, he knows his original Bell state before the entanglement swapping. That means, TP knows the correspondence of Bell states between particles $(C_{1}, B_{2})$ and particles $(B_{1}, A_{2})$, which is shown in one of Eqs. 2-5. Through TP's measurement outcome $M^{C}_{j}$ in Step 6, he can deduce Bob's corresponding result $M^{B}_{j}$. For example, suppose $M^{C^{'}}_{j}$ is $|\Phi^{-}\rangle$ and $M^{C}_{j}$ is $|\Psi^{+}\rangle$ , TP can deduce that $M^{B}_{j}$ is $|\Psi^{-}\rangle$ in terms of Eq. 4, and then gets $R^{B}_{j}=11$. Since $E^{B}_{j}(E^{B}_{j}=R^{B}_{j}\oplus G^{B}_{j})$ is sent through the classical channel, TP can get $E^{B}_{j}$ without being detected. By calculating $G^{B}_{j}=E^{B}_{j}\oplus R^{B}_{j}$, TP will obtain  $G^{B}_{j}$. For $j=1$ to $\lceil L/2\rceil$, TP can also get all of Bob's secret inputs $Y$.

\subsection{The Improvement}
To fix the loophole, we use the decoy photons instead of Bell states for eavesdropper checking. What is more, we introduce one-time eavesdropper checking instead of original twice ones in the Alice-TP and TP-Bob channels. To be specific, Step 3 and 5 in the original protocol need to be revised as follows.

Step $3^{*}$. Alice and TP prepare $L^{'}$ decoy photons randomly in four nonorthogonal photon states $\{|0\rangle, |1\rangle, |+\rangle, |-\rangle\}$ to form sequences $D_{A}$, $D_{C}$, and insert $D_{A}$, $D_{C}$ into $S^{A}_{2}$, $S^{C}_{2}$ at random positions to get $S^{A^{*}}_{2}$, $S^{C^{*}}_{2}$, respectively. Then they exchange sequences $S^{A^{*}}_{2}$, $S^{C^{*}}_{2}$. After receiving $S^{C^{*}}_{2}$, Alice utilizes sequence $D_{C}$ to perform the eavesdropper checking of TP-Alice quantum channel. The detailed procedures are as follows, (i) TP informs Alice the positions and the measurement bases (MBs) of the decoy photons. (ii) Alice performs the single-photon measurements on $D_{C}$ and publishes the outcomes. (iii) TP analyzes the error rate. If the error rate is higher than the threshold they preset, he will abort the protocol; otherwise, the TP-Alice quantum channel is secure, and Alice discards the checking particles and continues the next step. It should be noted that sequence $S^{A^{*}}_{2}$ remains unchanged on TP's hand.

Step $5^{*}$. Bob prepares $L^{'}$-length length nonorthogonal decoy photons sequence $D_{B}$, and randomly inserts $D_{B}$ into $S^{B}_{2}$ to get $S^{B^{*}}_{2}$, then he sends $S^{B^{*}}_{2}$ to TP. TP will perform the eavesdropper checking of the Bob-TP channel by using the same method as that in Step $3^{*}$. At the same time, TP sends the intact sequence $S^{A^{*}}_{2}$ to Bob. After receiving $S^{A^{*}}_{2}$, Bob will make one-time eavesdropper checking of the Alice-TP and TP-Bob channels with the help of Alice, the procedures are as follows. (i) Bob asks Alice to tell him the positions and MBs of sequence $D_{A}$. (ii) Bob uses the correct MBs to measure $D_{A}$, and publishes the outcomes. (iii)Alice analyzes the error rate, and ensures whether the Alice-TP and TP-Bob channels are secure or not. If the quantum channels are secure, Bob and TP discard the checking particles, and continue to the next step.

Let us examine the security of our improved protocol. Firstly, we will check whether our protocol can resist TP's measurement attack or not. Suppose TP wants to steal Alice's secrets. Alice and TP prepare decoy photons sequences $D_{A}$, $D_{C}$ in Step $3^{*}$, but Alice will not publish the positions of $D_{A}$ when TP receives the sequence $S^{A^{*}}_{2} (S^{A^{*}}_{2}=D_{A}||S^{A}_{2})$. Therefore, TP cannot properly extract $S^{A}_{2}$ from $S^{A^{*}}_{2}$ in Step 4, that is to say, he cannot obtain $M^{A}_{j}$ through making Bell-basis measurement. Of cause, TP cannot get Alice's secret inputs $X$. Suppose TP wants to steal Bob's secrets, and he will try to use the correspondence after the entanglement swapping of Bell states between particles $(C_{1}, B_{2})$ and particles $(B_{1}, A_{2})$ to deduce Bob's secret inputs. However, TP cannot know the Bell states of particles $(C_{1}, A_{2})$ before this entanglement swapping, so he can only guess $M^{A}_{j}$ as one of $\{|\Phi^{+}\rangle, |\Phi^{-}\rangle, |\Psi^{+}\rangle, |\Psi^{-}\rangle\}$ with the same probability (i.e., $1/4$). For instance, suppose the final measurement outcome of particles $(C_{1}, B_{2})$ is $|\Phi^{-}\rangle_{C_{1}B_{2}}$ in Step 6, TP tries to steal Bob's outcome in terms of Eqs 2-5. Since TP does not know the correct state of particles $(C_{1}, A_{2})$, he needs to take into account all the correspondences in these four equations. That is to say, TP can guess Bob's outcome to be $|\Phi^{-}\rangle_{B_{1}A_{2}}$, $|\Phi^{+}\rangle_{B_{1}A_{2}}$, $|\Psi^{-}\rangle_{B_{1}A_{2}}$ or $|\Psi^{+}\rangle_{B_{1}A_{2}}$ with the same probability $1/4$. As a result, TP cannot get any information about $R^{B}_{j}$, and then cannot obtain Bob's secrets through calculating $G^{B}_{j}=E^{B}_{j}\oplus R^{B}_{j}$. In addition, Liu et al.'s proved their protocol can resist the intercept-resend attack, the measure-resend attack and the entangle-measure attack in Ref. [25]. Since our improvement is similar to Liu et al.'s protocol, so it can also resist these attacks.

Since we use the decoy photons to make one-time eavesdropper checking instead of the original twice eavesdropper checking with Bell states, the particle efficiency has also been improved. For ensuring the checking photons is enough for performing eavesdropper checking, we suppose the number of checking particles is the same as that of message-encoded particles, that is, the length of checking particles $L^{'}=\lceil L/2\rceil$, and the length of secret inputs is $L=2\lceil L/2\rceil$. In QPCE, $\eta=\eta_{s}/\eta_{q}$ is usually used to calculate the particle efficiency, here $\eta_{s}$ represents the number of classical bits of secret inputs, and $\eta_{q}$ denotes the total number of particles used in the protocol. In Liu et al.'s protocol, TP, Alice and Bob totally generate $6\times \lceil L/2\rceil$ particles (i.e., $3\lceil L/2\rceil$ Bell states) to compare secret inputs ($X$ and $Y$), and $8\lceil L/2\rceil$ particles (i.e., 4$L^{'}$ Bell states) to make eavesdroppers checking, so $\eta_{s}=L$, $\eta_{q}=3L+4L=7L$, and then $\eta_{Liu}=L/7L\approx14.29\%$. In our improved version, totally $6\times \lceil L/2\rceil$ message-encoded particles (i.e., $3\lceil L/2\rceil$ Bell states) and $3\lceil L/2\rceil$ checking decoy photons are utilized, so $\eta_{Our}=L/4.5L\approx20.22\%$. Obviously, our particle efficiency is high than that of Liu et al.'s protocol. One point should be noticed, that is the particle efficiencies we calculated here are different from Ref. [18, 28], because they did not consider the photon consumption in the eavesdropper checking stage.

\section{Discussion and Conclusion}
The efficiency is one of the purposes and goals of our continuous improvement in quantum private computation. The eavesdropper checking is taken in every step of the particles' transmission in many QPCE protocols (including Liu et al.'s protocol), which makes these protocols anfractuous and inefficient. As we all know, the action of checking always requires the participant to have quantum devices, e.g., the qubit generating machine, the quantum memory, or the quantum measuring machine. Considering that these devices are expensive in the practical situation, it is unrealistic that every participant is equipped with all these quantum devices. Compared with this step-by-step checking, one-time checking may reduce the particles consumption and simplifies the protocol steps. On the other hand, since decoy photons are more economic and feasible in the practical application than other entangled states, Using decoy photons for eavesdropper checking may be a best choice.

In this paper, we find there is a security loophole in Liu et al.'s protocol, i.e., TP can launch the measurement attack to steal all the two participants' secret inputs. In order to fix the loophole, we utilize one-time eavesdropper checking with decoy photons instead of original twice eavesdropper checking with Bell states. And the improved protocol not only can resist TP's measurement attack, but also gets higher particle efficiency.

\ack
This work is supported by the National Nature Science Foundation of China (Grant Nos. 61103235, 61170321, 61373016 and 61373131), the Priority Academic Program Development of Jiangsu Higher Education Institutions (PAPD), the Natural Science Foundation of Jiangsu Province, China (BK2010570), and the Practice Inovation Trainng Program Projects for the Jiangsu College Students (201310300018Z).


\section*{References}
\begin{thebibliography}{28}
\bibitem{1} Bennett C H and Brassard G 1984 Quantum cryptography: public key distribution
and coin tossing \emph{Proc. of IEEE Int. Conf. on Computers, Systems and Signal Processing} 175-179.
\bibitem{2} Cleve R, Gottesman D and Lo H K 1999 \emph{Phys. Rev. Lett.} \textbf{83} 648-51 doi:10.1103/PhysRevLett.83.648
\bibitem{3} Xiao L, Long G L, Deng F G and Pan J W 2004 \emph{Phys. Rev.} A \textbf{69} 052307 doi:10.1103/PhysRevA.69.052307
\bibitem{4} Nie Y Y, Li Y H, Liu J C and Sang M H 2011 \emph{Opt. Commun.} \textbf{284} 1457-60 doi:10.1016/j.optcom.2010.10.084
\bibitem{5} Bennett C H 1992 \emph{Phys. Rev. Lett.} 68 3121-4
\bibitem{6} Yang J, Xu B J and Guo H 2012 \emph{Phys. Rev.} A \textbf{86} 042314 doi:10.1103/PhysRevA.86.042314
\bibitem{7} Barrett J, Colbeck R and Kent A 2013 \emph{Phys. Rev. Lett.} \textbf{110} 010503
\bibitem{8} Bouwmeester D, Pan J W, Mattle K, Eibl M, Weinfurter H and Zeilinger A 1997 Nature \textbf{390} 575-9 doi:10.1038/37539
\bibitem{9} Furusawa A, Sorensen J L, Braunstein S L, Fuchs C A, Kimble H J and Polzik E S 1998 Science \textbf{282} 706-9 doi:10.1126/science.282.5389.706
\bibitem{10} Deng F G, Long G L and Liu X S 2003 \emph{Phys. Rev.} A \textbf{68} 042317 doi:10.1103/PhysRevA.68.042317
\bibitem{11} Deng F G and Long G L 2004 \emph{Phys. Rev.} A \textbf{69} 052319 doi:10.1103/PhysRevA.69.052319
\bibitem{12} Liu W J, Chen H W, Li Z Q and Liu Z H 2008 \emph{Chin. Phys. Lett.} \textbf{25} 2354-7
\bibitem{13} Liu W J, Chen H W, Ma T H, Li Z Q, Liu Z H and Hu W B 2009 \emph{Chin. Phys.} B \textbf{18} 4105-9 doi:10.1088/1674-1056/18/10/007
\bibitem{14} Yao A C 1982 Protocols for secure computations \emph{Proc. 23rd Annual Symp. on Foundations of Computer Science} 160-164
\bibitem{15} BBoudot F, Schoenmakers B and Traore J 2001 \emph{Discrete Appl. Math.} \textbf{111} 23-36 doi:10.1016/s0166-218x(00)00342-5
\bibitem{16} Lo H K 1997 \emph{Phys. Rev.} A \textbf{56} 1154-62 doi:10.1103/PhysRevA.56.1154
\bibitem{17} Yang Y G and Wen Q Y 2009 \emph{J. Phys. A: Math. Theor.} \textbf{42} 055305 doi:10.1088/1751-8113/42/5/055305
\bibitem{18} Jia H Y, Wen Q Y, Li Y B and Gao F 2012 \emph{Int. J. Theor. Phys.} \textbf{51} 1187-94 doi:10.1007/s10773-011-0994-5
\bibitem{19} Liu W, Wang Y B, Jiang Z T and Cao Y Z 2012 \emph{Int. J. Theor. Phys.} \textbf{51} 69-77 doi:10.1007/s10773-011-0878-8
\bibitem{20} Huang W, Wen Q, Liu B, Gao F and Sun Y 2013 \emph{Sci. China Phys. Mech. Astron} \textbf{56} 1670-8 doi:10.1007/s11433-013-5224-0
\bibitem{21} Li Y B, Wang T Y, Chen H Y, Li M D and Yang Y T 2013 \emph{Int. J. Theor. Phys.} \textbf{52} 2818-25 doi:10.1007/s10773-013-1573-8
\bibitem{22} Sun Z W and Long D Y 2013 \emph{Int. J. Theor. Phys.} \textbf{52} 212-8 doi:10.1007/s10773-012-1321-5
\bibitem{23} Zi W, Guo F, Luo Y, Cao S and Wen Q 2013 \emph{Int. J. Theor. Phys.} \textbf{52} 3212-9 doi:10.1007/s10773-013-1616-1
\bibitem{24} Liu W J, Liu C, Liu Z H, Liu J F and Geng H T 2013 \emph{Int. J. Theor. Phys.} doi:10.1007/s10773-013-1807-9
\bibitem{25} Liu W, Wang Y B and Cui W 2012 \emph{Commun. Theor. Phys.} \textbf{57} 583-8 doi:10.1088/0253-6102/57/4/11
\bibitem{26} Liu, W.J., Liu, C., Wang, H.B., Jia, T.T., IETE Tech. Rev. (2013) (to be published).
\bibitem{27} Zhang W-W and Zhang K-J 2012 \emph{Quantum Inf. Process.} \textbf{12} 1981-90 doi:10.1007/s11128-012-0507-3
\bibitem{28} Tseng H Y, Lin J and Hwang T 2012 \emph{Quantum Inf. Process.} \textbf{11} 373-84 doi:10.1007/s11128-011-0251-0
\end{thebibliography}
\end{document}